# First-Principles Study of Large Gyrotropy in MnBi for Infrared Thermal Photonics

Md Roknuzzaman[1,2], Sathwik Bharadwaj[3], Yifan Wang[3], Chinmay Khandekar[3], Dan Jiao[3], Rajib Rahman[1], and Zubin Jacob[3, *]

[1]School of Physics, The University of New South Wales, Sydney, NSW 2052, Australia

[2]School of Chemistry, The University of Sydney, Sydney, NSW 2006, Australia

[3]Birck Nanotechnology Center, Elmore Family School of Electrical and Computer Engineering, Purdue University, West Lafayette, Indiana, 47907, USA

*zjacob@purdue.edu

**ABSTRACT**

Nonreciprocal gyrotropic materials have attracted significant interest recently in material physics, nanophotonics, and topological physics. Most of the well-known nonreciprocal materials, however, only show nonreciprocity under a strong external magnetic field and within a small segment of the electromagnetic spectrum. Here, through first-principles density functional theory calculations, we show that due to strong spin-orbit coupling manganese-bismuth (MnBi) exhibits nonreciprocity without any external magnetic field and a large gyrotropy in a broadband long-wavelength infrared regime (LWIR). Further, we design a multi-layer structure based on MnBi to obtain a maximum degree of spin-polarized thermal emission at 7 μm. The connection established here between large gyrotropy and the spin-polarized thermal emission points to a potential use of MnBi to develop spin-controlled thermal photonics platforms.

## I. INTRODUCTION

Thermal emission is a ubiquitous physical phenomenon in the environment. Any material heated to a non-zero temperature emits thermal photons. Fundamentally, thermal emission arises from the underlying fluctuating dipole moments inside the material [1,2]. Applications of near- and far-field thermal radiation include radiative cooling [3], thermal imaging [4], and energy harvesting [5]. In this regard, the search for infrared thermal photonic materials has gained significant interest. Magneto-optic materials [6,7], Weyl semimetals [8], topological insulators [9], and nanophotonic structures [10-13] have all been investigated recently to achieve control over the spectrum and the spin polarization of thermal radiation.

Discovery of materials with strong spin-angular momentum at infrared wavelengths will be useful to build polarized infrared light sources for thermal imaging polarimetry [14], night visions [15], and infrared chiral spectroscopy [16]. Nonreciprocal materials provide a pathway to achieve spin-polarized radiation sources [17]. Indium antimonide (InSb) subjected to an applied magnetic field is one of the most popular nonreciprocal media [18]. However, for practical applications, it is important to identify materials that exhibit nonreciprocity without any applied magnetic fields, especially in the long wavelength infrared (LWIR) regime. In this article, we propose manganese-bismuth (MnBi) as a material platform for spin-polarized thermal photonics, eliminating the requirement of an external magnetic field to achieve nonreciprocity.

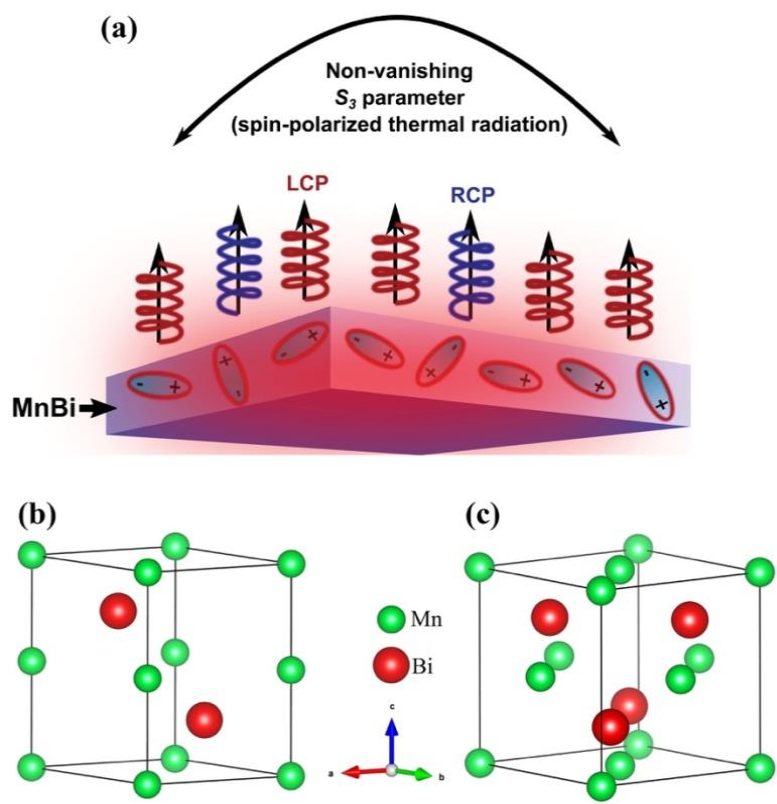

FIG. 1. (a) The schematic depicts the emission of spin-polarized thermal radiation originating from the intrinsic fluctuating dipole moments in a MnBi layer. Spin-polarization of the thermal radiation can be quantified by the non-vanishing Stoke parameter $S_3$. Crystal structures of the conventional unit cell of the low-temperature phases of MnBi: (b) Stable hexagonal phase and (c) Meta-stable cubic zinc-blende phase.

Manganese-bismuth has attracted significant attention due to its unique magneto-optic properties [19,20] and large magneto-crystalline anisotropy [21]. Most noticeable properties of MnBi include strong spin-orbit interaction [22], exceptionally large Kerr rotation [23], high coercivity which increases with temperature [24], and large uniaxial magnetic anisotropy [25]. Because of these extraordinary properties, MnBi has been used to fabricate magneto-optical memory devices and permanent magnets [25]. Furthermore, a recent study reveals a high anomalous Nernst effect (ANE) in MnBi which leads to a high thermopower and high thermoelectric conductivity [26]. Hence, MnBi is a promising platform for energy harvesting and cooling devices. The Curie temperature of MnBi is $T_c = 470$ K, significantly higher than room temperature [27]. Therefore, the material is a room-temperature permanent magnet [22]. Previous works have reported the electronic [19,22,28-42] as well as the optical and magneto-optic properties [19,30,32,34,42]. However, the nonreciprocity in MnBi for thermal photonics has not yet been investigated.

In this article, we show that a strong spin-orbit-coupling (SOC) in MnBi results in a large gyrotropy over a wide range of the LWIR regime. We perform first-principles density functional theory (DFT) calculations to obtain the electronic and optical properties of MnBi for its two possible low-temperature phases: hexagonal phase and zinc-blende phase. We identify that the origin of the strong SOC in MnBi is due to the half-filled 3*d* orbitals of the Mn atoms. We employ the nonreciprocal gyrotropy of MnBi to design a nanoscale spin-polarized radiation source based on a multilayer structure. The large gyrotropy in this material translates into a high degree of spin polarization of the emitted thermal radiation. Further, our results suggest that the hexagonal phase of MnBi displays a higher spin-polarized emissivity at IR wavelengths compared to its zinc-blende phase.

## II. COMPUTATIONAL DETAILS

### A. Methods

The first-principles Density Functional Theory (DFT) [43,44] calculations were performed within a Projector Augmented Wave (PAW) [45,46] framework by using the Vienna Ab initio Simulation Package (VASP) [47]. The generalized gradient approximation (GGA) of the Perdew–Burke–Ernzerhof (PBE) [48] functional was used to evaluate exchange-correlation energy. The conjugate-gradient algorithm [49] was used for the optimization of the crystal geometry by calculating forces and stress tensors, and by considering degrees-of-freedom in position, cell shape, and cell volume. Cut-off energy is taken to be at 350 eV for the expansion of the plane wave basis for both the hexagonal and zinc-blende phases. Monkhorst–Pack [50] $k$-point meshes of 13×13×8 and 8×8×8 were used for the sampling of the Brillouin Zone for the hexagonal and zinc-blende phases, respectively. The total energy convergence threshold of $10^{-6}$ eV and Gaussian smearing of 0.05 were used in all the calculations presented here.

At LWIR, a small variation in photon energy (or frequency) corresponds to a substantial change in photon wavelength. Therefore, extremely high grid points are required to explore the infrared region extensively, which is computationally expensive. We used a total grid point of hundred thousand for the calculation of the dielectric function, which corresponds to 190 data points per 1 eV interval. Additionally, SOC needs to be included in the calculations which is responsible for the large gyrotropy in MnBi in the absence of an applied magnetic field. This significantly increases the computational cost.

**B. Crystal Structure**

MnBi is known to exhibit two distinct phases: the ferromagnetic low-temperature phase (LTP) and the paramagnetic high-temperature phase (HTP) [22]. Most of the manganese alloys are antiferromagnetic because of the half-filled 3d orbitals of Mn, however, in contrast, MnBi is an exceptional ferromagnetic material [22]. The transition temperature from ferro to paramagnetic, and from para to ferromagnetic phases of MnBi were reported to be 628 K and 613 K, respectively [52]. The LTP of MnBi is found to crystalize in a hexagonal structure with the space group symmetry $p6\bar{3}/mmc$ (no. 194) [22]. The unit cell of hexagonal LTP MnBi contains two Mn atoms and two Bi atoms as shown in Fig. 1b. In the unit cell, the Mn atoms occupy 2a Wyckoff positions with fractional coordinates $(0,0,0)$ and $\left(0,0,\frac{1}{2}\right)$, and the Bi atoms occupy 2c Wyckoff positions with fractional coordinates $\left(\frac{1}{3},\frac{2}{3},\frac{1}{4}\right)$ and $\left(\frac{2}{3},\frac{2}{3},\frac{3}{4}\right)$. In addition to the stable low-temperature hexagonal phase, a meta-stable zincblende phase was also predicted to exist at low temperatures [51]. This zinc-blende phase was predicted to crystallize in a cubic structure with the space group $F\bar{4}3m$ (no. 216), and its crystal structure is the same as the sp-valent octet semiconductors such as GaAs, InAs, GaSb, InSb, and CdTe [51]. The unit cell of zinc-blende MnBi contains four Mn atoms and four Bi atoms as shown in Fig. 1c. In the unit cell, the Mn atoms occupy 4a Wyckoff positions with fractional coordinates $(0,0,0)$, $\left(0,\frac{1}{2},\frac{1}{2}\right)$, $\left(\frac{1}{2},0,\frac{1}{2}\right)$ and $\left(\frac{1}{2},\frac{1}{2},0\right)$, and the Bi atoms occupy 4c Wyckoff positions with fractional coordinates $\left(\frac{1}{4},\frac{1}{4},\frac{1}{4}\right)$, $\left(\frac{1}{4},\frac{3}{4},\frac{3}{4}\right)$, $\left(\frac{3}{4},\frac{1}{4},\frac{3}{4}\right)$ and $\left(\frac{3}{4},\frac{3}{4},\frac{1}{4}\right)$. We calculate the enthalpy of formation for both the considered structures. The calculated formation enthalpies of hexagonal and zinc-blende phases are $-4.3\ KJ/mole$ and $-3.7\ KJ/mole$, respectively. The more negative value of the formation enthalpy of hexagonal MnBi suggests that the formation of the hexagonal phase is more favourable than that of the zinc-blende phase during crystallization process.

**C. Calculations of LWIR Optical Constants**

The optical constants of a material can be derived from its real and imaginary parts of the dielectric function. The frequency-dependent complex dielectric function of crystalline solids is given by

$$\varepsilon(\omega) = \varepsilon_1(\omega) + i\varepsilon_2(\omega), \qquad (1)$$

where, $\varepsilon_1(\omega)$ and $\varepsilon_2(\omega)$ are the real and imaginary parts of the dielectric function, respectively. In cartesian coordinates, Eq. (1) can be expressed as

$$\varepsilon(\omega) = \begin{pmatrix} \varepsilon_{xx}(\omega) & \varepsilon_{xy}(\omega) & \varepsilon_{xz}(\omega) \\ \varepsilon_{yx}(\omega) & \varepsilon_{yy}(\omega) & \varepsilon_{yz}(\omega) \\ \varepsilon_{zx}(\omega) & \varepsilon_{zy}(\omega) & \varepsilon_{zz}(\omega) \end{pmatrix}. \quad (2)$$

At LWIR wavelengths, there is a significant contribution from intra-band transitions in addition to inter-band transitions. Therefore, the dielectric function is a sum of inter- and intra-band contributions, $\varepsilon^{inter}(\omega)$ and $\varepsilon^{intra}(\omega)$ respectively, given by

$$\varepsilon(\omega) = \varepsilon^{inter}(\omega) + \varepsilon^{intra}(\omega), \quad (3)$$

The inter-band contribution to the dielectric function can be derived from the first-order time-dependent perturbation theory [53], and the corresponding expression is given by

$$\varepsilon_{\alpha\beta}^{inter}(\omega) = 1 - \frac{8\pi e^2}{\Omega} \lim_{\substack{q \to 0 \\ \alpha \to 0}} \frac{1}{q^2} \sum_{k,v,c} \frac{\langle \psi_{k+qe_\alpha}^c | e^{iq\cdot r} | \psi_k^v \rangle \langle \psi_k^v | e^{-iq\cdot r} | \psi_{k+qe_\beta}^c \rangle}{(E_{k+q}^c - E_k^v - \hbar\omega - i\hbar\alpha)} + cc, (4)$$

where, $\omega$ is the phonon frequency, $e$ is the charge of an electron, $\Omega$ is the volume of a unit cell, $\boldsymbol{q}$ is the photon momentum, $\boldsymbol{r}$ is the radius vector, and $\psi_{k+q}^c$ and $\psi_k^v$ are the wavefunctions for conduction and valence band electrons, respectively at a given electron wavevector $\boldsymbol{k}$. In practice, we evaluate the imaginary part of the dielectric function numerically and calculate the real part by using the Kramers-Kronig relation, given by

$$\varepsilon_1^{inter}(\omega) = 1 + \frac{2}{\pi} P \int_0^\infty \frac{\omega' \varepsilon_2^{inter}(\omega') d\omega'}{(\omega'^2 - \omega^2)}. \quad (5)$$

Further, a free-electron plasma model is used to calculate the intra-band contributions to the dielectric function, given by

$$\varepsilon^{intra}(\omega) = 1 - \frac{\omega_p^2}{\omega(\omega + i\gamma)}, \quad (6)$$

where, the plasma frequency $\omega_p$ can be obtained from first-principles calculations. The inverse lifetime $\gamma$ can have a value between 0 and 1 eV [54]. The complex optical conductivity $\sigma(\omega)$ is determined through the relation,

$$\sigma(\omega) = -i\frac{\omega}{4\pi}[\varepsilon(\omega) - 1]. \quad (7)$$

The hexagonal MnBi crystal has a tetragonal symmetry with the polar Kerr magnetization geometry. Therefore, both the fourfold axes and the magnetization are perpendicular to the surface of the sample, and the z-axis is chosen to be parallel to these. In this case, the dielectric tensor has only three independent components (diagonal $\varepsilon_{xx}$ and $\varepsilon_{zz}$, and off-diagonal $\varepsilon_{xy}$) and can be represented as

$$\varepsilon(\omega) = \begin{pmatrix} \varepsilon_{xx} & \varepsilon_{xy} & 0 \\ -\varepsilon_{xy} & \varepsilon_{xx} & 0 \\ 0 & 0 & \varepsilon_{zz} \end{pmatrix}. \quad (8)$$

Similarly, the conductivity tensor for hexagonal MnBi has the form [19]

$$\sigma(\omega) = \begin{pmatrix} \sigma_{xx} & \sigma_{xy} & 0 \\ -\sigma_{xy} & \sigma_{xx} & 0 \\ 0 & 0 & \sigma_{zz} \end{pmatrix}. \quad (9)$$

On the other hand, the zinc-blende MnBi has a cubic crystallographic structure, and its dielectric tensor has only two independent components (diagonal $\varepsilon_{xx}$ and off-diagonal $\varepsilon_{xy}$) and can be represented as

$$\varepsilon(\omega) = \begin{pmatrix} \varepsilon_{xx} & \varepsilon_{xy} & \varepsilon_{xy} \\ -\varepsilon_{xy} & \varepsilon_{xx} & \varepsilon_{xy} \\ -\varepsilon_{xy} & -\varepsilon_{xy} & \varepsilon_{xx} \end{pmatrix}. \quad (10)$$

Similarly, the conductivity tensor of zinc-blende MnBi has two independent components: $\sigma_{xx}$ and $\sigma_{xy}$.

## III. RESULTS AND DISCUSSION

### A. Electronic Properties

Herein, we present the electronic band structure obtained using the Perdew–Burke–Ernzerhof (PBE) method including spin polarization and spin-orbit-coupling (SOC). We further investigate the total and projected densities of states (PDOS) to understand the individual atomic orbital contributions to the total density of states (DOS). In this work, the Fermi level is always considered at 0 eV. In Figs. 2a and 2b we have plotted the spin-polarized electronic band structure for hexagonal and zinc-blende MnBi, respectively. It is evident from Fig. 2a that the bands for both up-spin and down-spin electrons cross the Fermi level, signifying that the hexagonal MnBi is metallic for both up- and down-spin electrons. On the other hand, the spin-polarized band structure of zinc-blende MnBi (Fig. 2b) reveals that there is a bandgap for down-spin electrons, but no bandgap for the up-spin electrons. Therefore, the zinc-blende MnBi is a half-metal [51].

We also investigate the effect of SOC on the band structures of hexagonal and zinc-blende MnBi shown in Fig. 2c and 2d, respectively. The inclusion of the SOC significantly affects the band structures for both the hexagonal and zinc-blende phases as a notable deviation is observed in the evolution of bands when compared to the calculations performed without including SOC. Also, the effect of SOC is greater for hexagonal MnBi while compared to zinc-blende MnBi as the band deviation near the Fermi level is more prominent for the hexagonal phase with the inclusion of SOC.

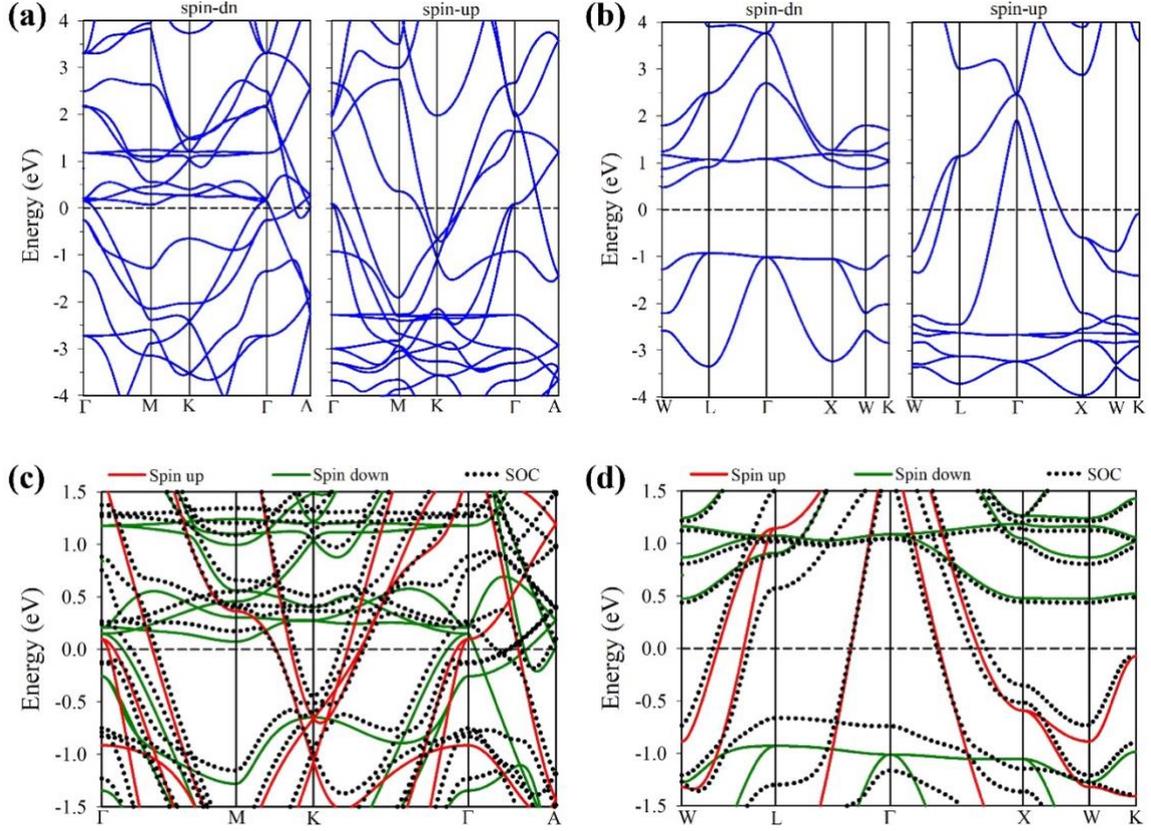

FIG. 2. Electronic band structures of hexagonal and zinc-blende phases of MnBi. Spin-polarized band structure for (a) hexagonal and (b) zinc-blende MnBi. Effect of SOC on the band structure for (c) hexagonal and (d) zinc-blende MnBi. We observe that the inclusion of spin-orbit coupling leads to significant changes in the band structures for both the hexagonal and zinc-blende phases. The horizontal dashed lines here represent the band structure obtained with the inclusion of SOC.

In Fig. 3, we investigate the total density of states (TDOS) for both the hexagonal and zinc-blende MnBi. We further resolve TDOS into PDOS to understand the individual contribution from each specific orbital. Results suggest that the 6*p* orbital of bismuth has partial contribution to the valence band at lower energies (-2 to -1 eV). However, the contribution near the Fermi level is mainly due to the 3*d* orbital of the Mn atom. The rest of the orbitals of Mn and Bi have negligible contributions to the TDOS. The 3*d* orbital of Mn has 5 electrons, *i.e.*, the orbital is half filled. Interestingly, the half (or, partially) filled *d* and *f* orbital is a major factor for the observed magnetic properties of a material. Therefore, the strong SOC in MnBi is due to the half-filled 3*d* orbital of Mn as evident from the TDOS at the Fermi level.

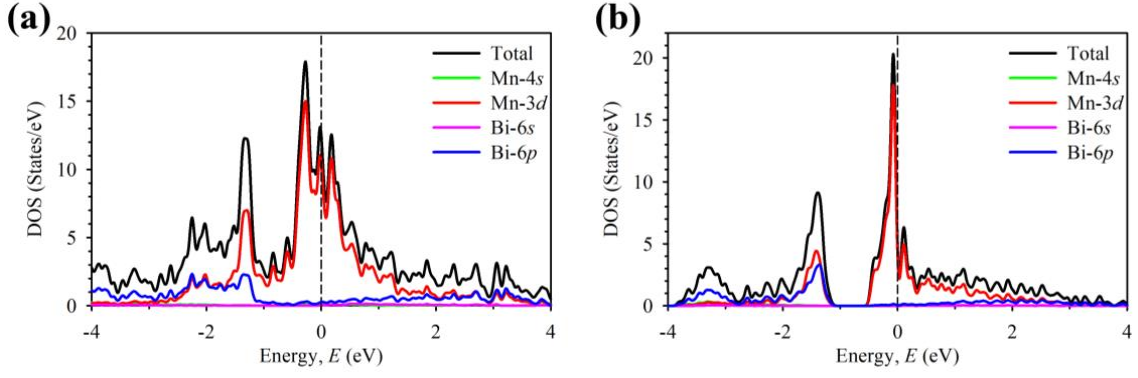

FIG. 3. The calculated total and projected density of states for (a) hexagonal and (b) zinc-blende MnBi. It is found from the projected density of states that the 3*d*-orbital of Mn is the major contributor to the total DOS at the Fermi level. Also, the projected densities of states for Mn-4*s*, Bi-6*s* and Bi-6*p* at around the Fermi level are nearly zero. Hence, the strong SOC found in MnBi is due to half-filled 3*d* orbital of the Mn atom.

**B. LWIR Non-Reciprocal Optical Properties**

In this section, we study the frequency-dependent dielectric function and conductivity characteristics. For a metallic system like MnBi, the intra-band component has a considerable contribution to the total dielectric function. Therefore, we consider both the inter-band and intra-band components to compute the total dielectric function. Also, Eq. (6) signify that the value of inverse lifetime $\gamma$ is crucial for the calculation of intra-band dielectric function in the low energy regime, especially at IR wavelengths. Herein, we treat MnBi as an ordinary metallic system and consider the typical experimental value of $\gamma$ for metal as 0.1 eV [54], whereas good conductors like silver and gold have $\gamma$ of 0.02 eV and 0.07 eV, respectively [55,56]. Furthermore, we calculate the full plasma frequency ($\omega_p$) tensor and use this to estimate the intra-band contributions to the dielectric function. The calculated values of $\omega_p$ for hexagonal MnBi are 1.66, 3.79 and 4.42 eV for tensor elements xy, xx and zz, respectively. On the other hand, the calculated values of $\omega_p$ for zinc-blende MnBi are found to be 1.65 and 3.98 eV for xy and xx, respectively.

The calculated imaginary and real parts of the dielectric function are plotted in Fig. 4 for infrared (IR) and visible regions (inset) of the electromagnetic spectra. We consider a wide spectrum of wavelengths varying from 5 to 20 μm to capture the optical properties in the technologically important LWIR regime. The off-diagonal and the diagonal components of the imaginary part of the dielectric function for both the hexagonal and zinc-blende MnBi are presented in Fig. 4a. At 5 μm, $Im(\varepsilon_{xy})$ has a value of ~ 40 for both structures, and the value is seen to increase gradually with increasing wavelength. The value of $Im(\varepsilon_{xy})$ at 20 μm is as high as ~ 400 for the hexagonal phase. On the other hand, at 20 μm, $Im(\varepsilon_{xx})$ for hexagonal and zinc-blende phases are ~ 2,000 and ~ 2,400, respectively, and $Im(\varepsilon_{zz})$ for the hexagonal phase is ~ 2,400. Therefore, for the imaginary part of the dielectric function, the off-diagonal component is 5 to 6 times smaller than the diagonal components at higher wavelengths. Moreover, $Im(\varepsilon_{xy})$ is higher for the hexagonal phase than the zinc-blende phase for the entire range of considered IR wavelengths. However, $Im(\varepsilon_{xy})$ for zinc-blende MnBi is found to be higher than that of the hexagonal phase for visible wavelengths.

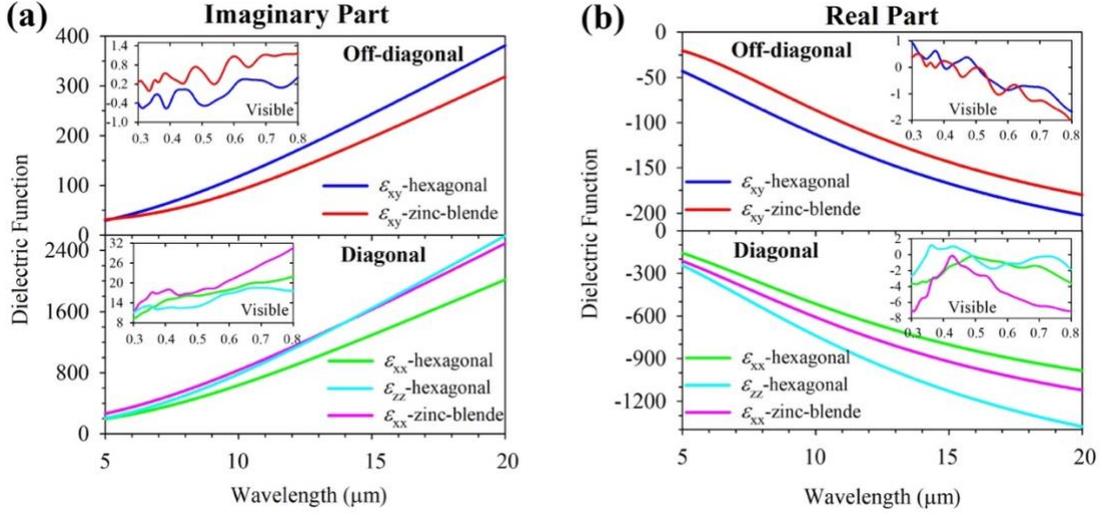

FIG. 4. LWIR dielectric function for hexagonal and zinc-blende phases of MnBi. (a) Independent components of the imaginary part of the dielectric tensor due to off-diagonal and diagonal elements. (b) Independent components of the real part of the dielectric tensor due to off-diagonal and diagonal elements.

The off-diagonal and the diagonal components of the real part of the dielectric function are investigated and plotted in Fig. 4b. At low energy, both the off-diagonal and diagonal components of the real dielectric function have negative values signifying that the considerable contributions from intra-band transitions at LWIR. The $Re(\varepsilon_{xy})$ is more negative for the hexagonal phase while compared to that of the zinc-blende phase in the IR regime. However, $Re(\varepsilon_{xx})$ is less negative in the hexagonal phase. Overall, our calculations reveal that the imaginary off-diagonal dielectric function of MnBi has a significant value compared to its real part, especially at IR wavelengths. This leads to a large gyrotropy without any applied magnetic field which is extremely useful for thermal spin photonics.

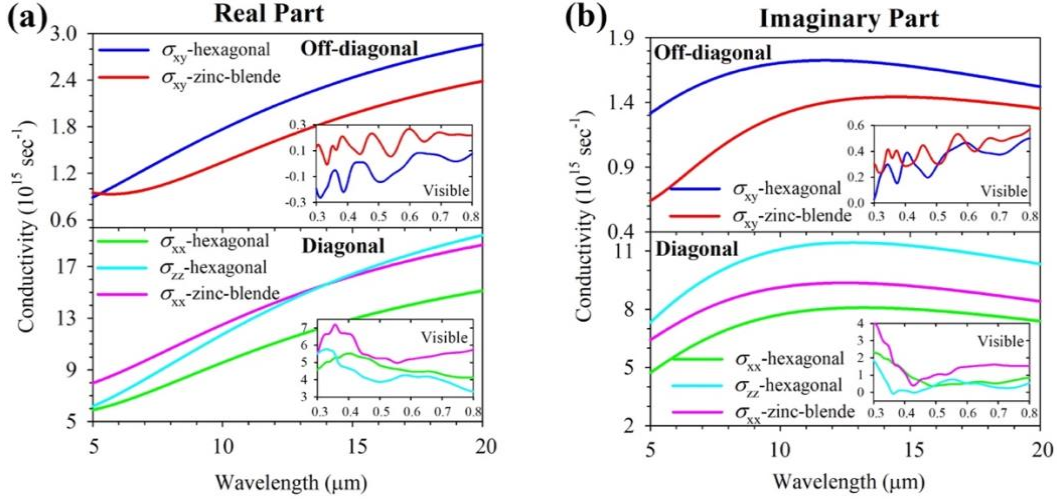

FIG. 5. LWIR optical conductivity of hexagonal and zinc-blende phases of MnBi. (a) Independent components of the real part of the conductivity tensor due to off-diagonal and diagonal parts. (b) Independent components of the imaginary part of the dielectric tensor due to off-diagonal and diagonal parts.

We further calculate the full optical conductivity tensor for both the hexagonal and zinc-blende phases of MnBi. The calculated independent components of the real and imaginary parts of the conductivity tensors for the incident IR and visible region are presented in Fig. 5. Like the imaginary dielectric function, the off-diagonal real part of the conductivity has significant values at IR wavelengths. Also, the off-diagonal real part of the conductivity of the hexagonal phase is higher than that of the zinc-blende phase. Overall, both the off-diagonal and diagonal components of the real part of the conductivity increase with increasing wavelength. On the other hand, the imaginary part of the conductivity is observed to have a small variation over the considered IR range.

## C. Gyrotropy and Infrared Thermal Photonics

The gyrotropy of a material is defined as the ratio of $|Im(\varepsilon_{xy})|$ and $|Re(\varepsilon_{xx})|$, given by

$$g = \frac{|Im(\varepsilon_{xy})|}{|Re(\varepsilon_{xx})|}. \qquad (11)$$

In Fig. 6(a) and 6(b), we plot the gyrotropy for the hexagonal and zinc-blende phases of MnBi as a function of wavelength in the LWIR regime. We observe that the hexagonal phase of MnBi displays higher gyrotropy while compared to that of the zinc-blende phase. In the next section, we employ the gyrotropy of MnBi in the IR region and show its potential application in realizing a nanoscale spin-polarized radiation source.

We consider a semi-infinite half space of multi-layer design of MnBi at temperature $T = 300\ K$ emitting thermal radiation into the vacuum half space (environment) at $T_0 = 0\ K$. We evaluate the thermal emission in spherical coordinates $(\theta, \phi)$. The thermal radiation power emitted per unit wavelength $d\lambda$ per unit solid angle $d\Omega$ per unit surface area $dA$ for a given polarization state $\hat{e}$ is given by,

$$P_{rad}(\theta,\phi,\lambda,\hat{e}) = \eta(\theta,\phi,\lambda,\hat{e})\frac{I_{bb}(\lambda,T)}{2}cos(\theta)d\lambda\, d\Omega\, dA, \qquad (12)$$

where $\eta$ is the dimensionless emissivity $\in [0, 1]$, $I_{bb}(\lambda,T) = \frac{2hc^2}{\lambda^5}\frac{1}{e^{hc/(\lambda k_B T)}-1}$ is the Planck distribution function for a blackbody at temperature $T$, $h$ is the Planck's constant and $k_B$ is the Boltzmann constant. A factor of 2 in the denominator of Eq. (12) accounts for the two orthogonal polarization states.

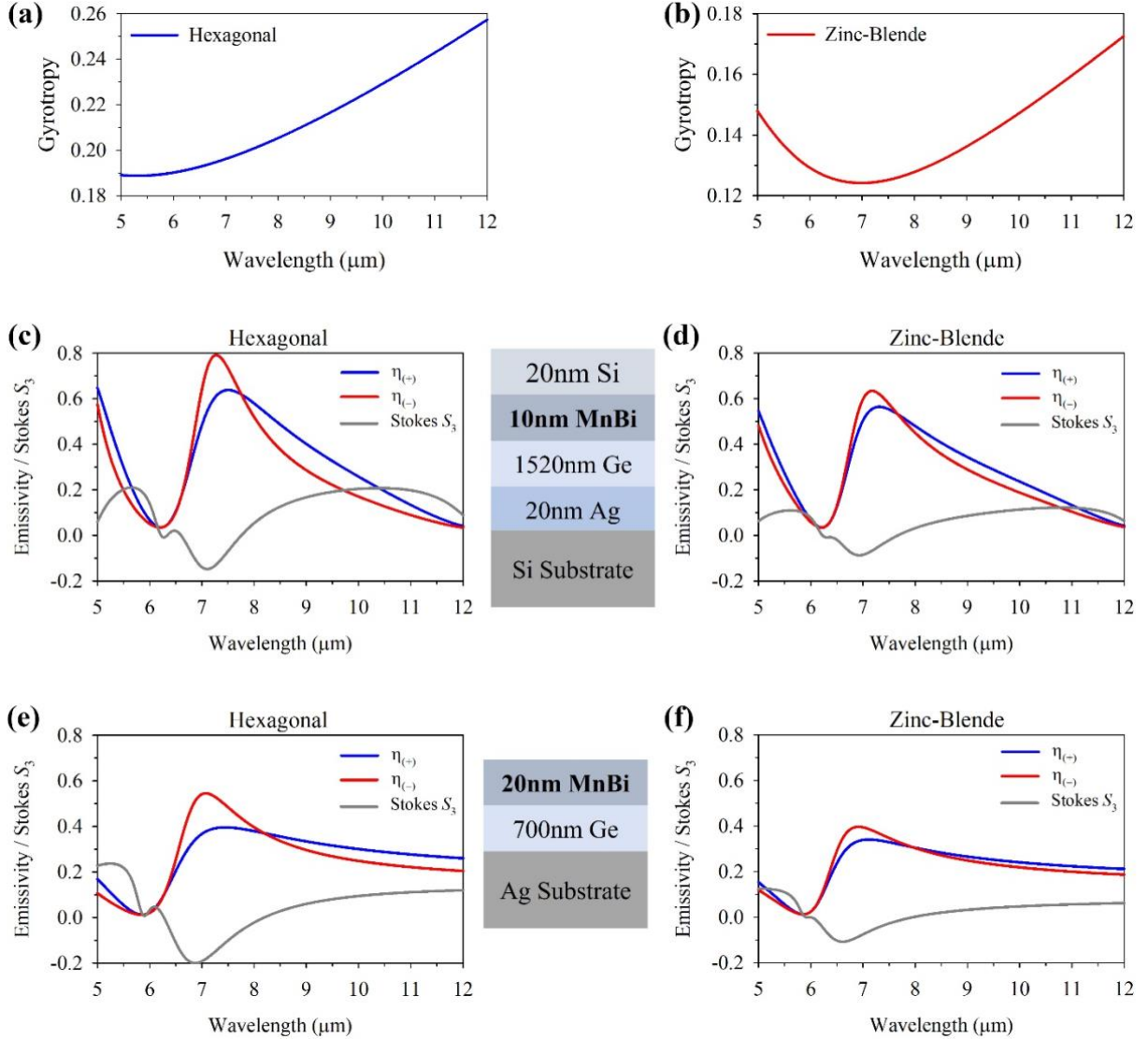

FIG. 6. The gyrotropy ratio calculated from the DFT calculations, and the simulated circularly polarized (CP) thermal emission for hexagonal and zinc-blende MnBi in the LWIR regime. Gyrotropy of (a) hexagonal and (b) zinc-blende MnBi. Emissivity for right circularly polarized (RCP) denoted as $\eta_{(+)}$ and left circularly polarized (LCP) denoted as $\eta_{(-)}$ as well as Stokes parameter ($S_3$) for (c, e) hexagonal and (d, f) zinc-blende MnBi for two different device geometries. Observed asymmetry in $\eta_{(+)}$ and $\eta_{(-)}$ translates into a spin-polarized thermal emission in MnBi with maximum spin polarization at 7 μm.

Equation (12) for $P_{rad}(\theta, \phi, \lambda, \hat{e})$ has been derived within the radiometry framework using a detailed balance of energy and momenta or within the scattering formulation of fluctuational electrodynamics [2]. Since we are interested in the spin-polarization of light, we calculate the polarization-dependent emissivity in the eigen basis of right circularly polarized (RCP) and left circularly polarized (LCP) states. We can define the degree of circular polarization, i.e., the 3rd Stokes parameters, for the thermal emission as

$$S_3(\theta, \phi, \lambda) = \frac{\eta_{(+)} - \eta_{(-)}}{\eta_{(+)} + \eta_{(-)}}, \tag{13}$$

where $\eta_{(\pm)}$ denotes the emissivity of RCP and LCP, respectively. For a planar geometry, the emission direction for thermal radiation is given by the angles $\theta, \phi$ for the propagation wavevector $\hat{k}$. Eigenvectors of the associated transverse electric (s), and transverse magnetic (p) polarization for the plane of incidence spanned by $\hat{k}$ and $\hat{z}$ (normal to the slab surface) are given by

$$\hat{k} = \begin{bmatrix} \sin\theta\cos\phi \\ \sin\theta\sin\phi \\ \cos\theta \end{bmatrix}, \hat{e}_s = \begin{bmatrix} +\sin\phi \\ -\cos\phi \\ 0 \end{bmatrix}, \hat{e}_p = \hat{e}_s \times \hat{k} = \begin{bmatrix} -\cos\theta\cos\phi \\ -\cos\theta\sin\phi \\ \sin\theta \end{bmatrix}. \tag{14}$$

The RCP and LCP eigenvector corresponds to $\hat{e}_s \pm i\hat{e}_p$, respectively. Hence, the spin angular momentum for a photon along the propagation direction is $\pm\hbar$ for RCP and LCP, respectively. The emissivity of RCP and LCP photons in terms of reflectance is given by

$$\eta_{(+)}(\omega, \theta, \phi) = 1 - R_{(++)}(\omega, \theta, \phi + \pi) - R_{(+-)}(\omega, \theta, \phi + \pi),$$
$$\eta_{(-)}(\omega, \theta, \phi) = 1 - R_{(--)}(\omega, \theta, \phi + \pi) - R_{(-+)}(\omega, \theta, \phi + \pi), \tag{15}$$

where, $R_{(ij)}(\omega, \theta, \phi)$ for $i, j \in \{+, -\}$ denotes the polarization interconversion reflectance for the light of angular frequency ω incident in the direction characterized by the angles (θ, φ). These reflectance coefficients depend on the associated Fresnel reflection coefficients in $\hat{e}_s, \hat{e}_p$ basis and given by

$$R_{(++/--)} = |(r_{ss} + r_{pp}) \pm i(r_{sp} - r_{ps})|^2/4,$$
$$R_{(-+/+-)} = |(r_{ss} - r_{pp}) \pm i(r_{sp} + r_{ps})|^2/4. \tag{16}$$

where $r_{jk}(\omega, \theta, \phi)$ denotes the amplitude of j-polarized reflected light due to incident k-polarized light of unit amplitude with frequency ω, and for brevity we have omitted $(\omega, \theta, \phi)$-dependence on either side of the expression. These reflection coefficients can be evaluated by solving the boundary conditions. Numerical codes to obtain these coefficients for a general bi-anisotropic media are made available on GitHub (https://github.com/chinmayCK/Fresnel).

In Fig. 6, we plot the emissivity of RCP and LCP, and the Stoke parameter $S_3$ as a function of LWIR wavelength in two designs of stratified materials. Both designs utilize a thin layer of MnBi to generate circularly polarized thermal emission. We optimize the design to ensure the emissivity and $S_3$ has a center around λ = 7 μm. LWIR optical properties in MnBi are dominated by intra-band contributions. The Ge layer between reflective metallic MnBi and Ag on its two sides acts as a Fabry Perot cavity, which resonantly enhances the strength of nonreciprocity. Hence, we design a sandwiched structure of thin layer MnBi with Ag to enhance $S_3$. For both hexagonal MnBi and zinc-Blende MnBi, our calculations provide the connection between strong gyrotropy and large degree of CP thermal radiation. For MnBi thin film on top of a Ge sandwich layer, grown on an Ag substrate, we observe that $S_3$ can be tuned to reach 0.2 at 7 μm. Hence, the large gyrotropy of MnBi can be employed to design a nanoscale spin radiation source for LWIR applications.

**IV. CONCLUSIONS**

In summary, a first-principles DFT study has been performed to investigate the electronic, and LWIR optical properties of MnBi. Strong spin-orbit coupling in MnBi induces a large gyrotropy even in the absence of an external magnetic field. We observed that the SOC in MnBi is due to the half-filled 3*d* orbital of Mn atoms. Furthermore, the thermal emission in MnBi for a multi-layer design has been studied for right circularly polarized (RCP) and left circularly polarized (LCP) photon emission. A significant degree of spin polarization of the emissivity is observed due to the large gyrotropy of MnBi. The given result suggests that hexagonal MnBi has a better performance for spin-polarized emissivity at LWIR compared to zinc-blende MnBi. Hence, hexagonal MnBi is potentially a technologically important candidate to form a spin-polarized radiation source for several infrared thermal photonics applications.

**ACKNOWLEDGEMENT**

This work was supported by the United States Department of Energy, Office of Basic Sciences under DE-SC0017717, and the Defence Advanced Research Projects Agency (DARPA) under the Nascent Light-Matter Interactions (NLM) program. Calculations presented here were conducted by using the computational resources from National Computational Infrastructure (NCI) under an NCMAS 2021 allocation, supported by the Australian Government.